\documentclass[aps,prd,fleqn,superscriptaddress]{revtex4}
\usepackage{graphicx,color,natbib,float}
\usepackage{amsmath,amssymb,amsfonts}
\newcommand{\bse}{\begin{subequations}}
\newcommand{\ese}{\end{subequations}}
\newcommand{\be}{\begin{equation}}
\newcommand{\ee}{\end{equation}}
\newcommand{\bea}{\begin{eqnarray}}
\newcommand{\eea}{\end{eqnarray}}
\newcommand{\ba}{\begin{array}}
\newcommand{\ea}{\end{array}}

\input amssym.def
\input amssym.tex

\usepackage[colorlinks=true, linkcolor=blue, bookmarks=true]{hyperref}
\begin{document}
\hfill%
\vbox{
\halign{#\hfil \cr
IPM/P-2020/012\cr
}
}
\vspace{1cm}
\title{On Volume Subregion Complexity in  Non-Conformal Theories}
\author{M. Asadi}
\email{{\rm{m}}$_{}$ asadi@ipm.ir}
\affiliation{School of Physics, Institute for Research in Fundamental Sciences (IPM),
P.O.Box 19395-5531, Tehran, Iran}
\begin{abstract}
We study the volume prescription of  the holographic subregion complexity in a holographic $5-$dimensional model consisting of Einstein gravity coupled to a scalar field with a non-trivial potential. The dual $4-$dimensional gauge theory is not conformal and exhibits a RG flow between two different fixed points. In both zero and finite temperature we show that the holographic subregion complexity can be used as a measure of non-conformality of the  model. This quantity exhibits also a monotonic behaviour in terms of the size of the entangling region, like the behaviour of the entanglement entropy in this setup. There is also a finite jump due to the disentangling transition between connected and disconnected minimal surfaces for holographic renormalized subregion complexity at zero temperature. 
\end{abstract}

\maketitle

\tableofcontents
\section{Introduction}
The gauge/gravity duality is a conjectured relationship between quantum field theory and gravity. The underlying duality provides an important framework to study key properties of the boundary field theory dual to some gravitational theory on the bulk side \cite{Maldacena}. The most significant example of gauge/gravity duality is the AdS/CFT correspondence which proposes a duality between asymptotically AdS spacetimes in $d+1$ dimensions and $d-$dimensional conformal field theories. This correspondence also indicates that there could be a deep relation between quantum gravity and quantum information theory, in the sense that there could be a holographic dual for some quantum information theory objects. Therefore, one could expect  that the nature of spacetime geometry could be understood from quantum information theory. This framework has been applied to study quantities such as entanglement entropy, n-partite information and recently extended to the quantum computational complexity in field theory. The generalization of gauge/gravity duality to field theories which are not conformal seems to be important. It is then interesting to develop our understanding of this duality for more general cases. There are many different families of non-conformal theories which one can study the effect of the non-coformality on their physical quantities \cite{Attems:2016ugt,Pang:2015lka}.
 
The entanglement entropy is a measure of the quantum correlations of a quantum state which is extremely useful in many quantum systems, ranging from condensed matter physics to black hole physics. The entanglement entropy has a  holographic dual given by the area of minimal surface extended on the bulk whose boundary coincides with the boundary of the subregion \cite{Ryu:2006bv,Ryu:2006ef}.

Aside from the entanglement entropy, another information theoretic quantity which is receiving a great attention is the quantum complexity. Quantum complexity describes how many simple elementary gates are needed to obtain a particular state from some chosen reference state, for the review see\cite{Aaronson:2016vto}. From the AdS/CFT correspondence, there have been two different proposals on holographic complexity, which are referred to as the complexity= action (CA) conjecture \cite{Stanford:2014jda} and the complexity=volume (CV) conjecture \cite{Brown:2015bva,Brown:2015lvg}, respectively. The first conjecture relates quantum complexity to the size of the wormhole. The linear growth of complexity in a thermalizing system is then holographically dual to the linear growth of the wormhole in an AdS black hole geometry. The second conjecture states that the complexity is given by the bulk action evaluated on the Wheeler-deWitt patch attached at some boundary time $t$.  There are by now a large number of papers developing and extending these ideas \cite{Alishahiha:2018lfv}. 

There has been much interest to generalize the notion of holographic complexity to subregions. That is, one would like to evaluate the complexity of the mixed state produced by reducing the boundary state to a specific subregion of the boundary time slice. Holographic subregion complexity has recently been studied in a variety of works with several proposals analogous to CV and CA complexity proposals, for both time-independent and time-dependent geometries \cite{Alishahiha:2015rta, Ben-Ami:2016qex, Carmi:2016wjl, Ling:2018xpc, Caceres:2018luq}. In \cite{Alishahiha:2015rta} the CV proposal has been generalized to the subregion complexity for time-independent geometries. Indeed, for a static bulk geometry, the CV duality for subregions evaluates the volume of the extremal codimension-one surface in the bulk which is bounded by the subregion on the asymptotic boundary and the Ryu-Takayanagi (RT) surface  for this subregion.
 
The background we have considered in this paper is a holographic $5$-dimensional model consisting of Einstein gravity coupled to a scalar field with a non-trivial potential, which is negative and has a minimum and a maximum for finite values of scalar field. Each of these extrema corresponds to an AdS$_5$ solution with different radii\cite{Attems:2016ugt}. In the gauge theory the $4$-dimensional  boundary  is not conformal and, at zero temperature,  flows from an UV fixed point to an IR fixed point. On the gravity side, this renormalization group is dual  to a geometry that interpolates between two AdS spaces. We are now interested in studying subregion CV proposal  in the non-conformal theories and study the effect of field theory parameters such as energy scale and model parameter on it.
\section{Review on the background}
The holographic model  we study here  is a five-dimensional Einstein gravity coupled to a scalar field with a non-trivial potential whose action is given by
\begin{eqnarray}\label{action}
S=\frac{2}{(G_N^{5})^2}\int d^5x\:\sqrt{-g}\:\big[\frac{1}{4}\mathcal{R}\:-\:\frac{1}{2}(\bigtriangledown\phi)^2\:-\:V(\phi)\big],
\end{eqnarray}
where $G_N^{5}$ is the five-dimensional Newton constant and $\mathcal{R}$ is the Ricci scalar of curvature corresponding to the metric $g$. Scalar field and its potential are also denoted by $\phi$ and $V(\phi)$, respectively. In order to have a bottom-up model  the following  potential has been chosen \cite{Attems:2016ugt}
 
\begin{align}
\begin{split}
L^2V(\phi)&=-3-\frac{3}{2}\phi^2 -\frac{1}{3}\phi^4 + \big(\frac{1}{3\phi_M^2 }+\frac{1}{2\phi_M^4}\big)\phi^6\\
& -\frac{1}{12\phi_M^4}\phi^8 .
\end{split}
\end{align}

This potential possess a maximum at $\phi=0$ and a minimum at $\phi=\phi_M>0$, each of them corresponds to an  $AdS_5$ background with different radii $L^2=\frac{-3}{V(\phi)}$. Each of these extremal is dual to a fixed point of the $RG$, with number of degrees of freedom $N^2\propto L^3$,  from the UV fixed point at $\phi=0$ to the IR fixed point at $\phi=\phi_M>0$. 
%The radii of these asymptotically $AdS_5$ take the following form \cite{Attems:2016ugt}
%\begin{eqnarray}\label{sAB}
%{L_{UV(IR)}=\sqrt{\frac{-3}{V(\phi)}}}=
%\begin{cases}
%L_{UV}=L & \: \phi=0 , \\
%L_{IR}=\frac{L}{1+\frac{\phi_M ^2}{6}} & \: \phi=\phi_M . \\
%\end{cases}
%\end{eqnarray}
%According to \eqref{sAB}, it is clearly seen that $L_{IR}<L_{UV}=L$ and then  the smaller number of degrees of freedom live in the IR limit, $i.e$. $N_{IR}<N_{UV}$. Furtheremore, as one  increases $\phi_M$ the difference in degrees of freedom between the UV and IR  fixed points increases. If one interested in domain-wall solutions which are interpolating between the two underlying $AdS_5$ backgrounds, the vacuum solutions to the Einstein equations can be obtained from the action (\ref{action}) .
 The parameterized metric for arbitrary $\phi_M$ can be read
 \begin{eqnarray}\label{metric}
ds^2=e^{2A(r)}(-dt^2+d \vec{x} ^2)+dr^2,
\end{eqnarray}
where 
\begin{align}
e^{2A(r)}&=\frac{\Lambda ^2 L^2}{\phi ^2}(1-\frac{\phi ^2}{\phi _M ^2})^{1+\frac{\phi _M^2}{6}}\,\,\,\,\, e^{-\frac{\phi ^2}{6}},\\
\phi (r)&=\frac{\Lambda L e^{-\frac{r}{L}}}{\sqrt{1+\frac{\Lambda ^2L^2}{\phi _M^2}e^{-\frac{2r}{L}}}},
\end{align}
where $\Lambda$  is the energy scale that break the conformal symmetry in the dual gauge theory.
% It is also related to the asymptotic value of the scalar field, $i.e$. $\phi(r\rightarrow \infty)$ and $\phi _M$ is the model parameter.

 The thermal physics of the non-conformal model described in \eqref{action} is given by the following geometry 
 \begin{eqnarray}\label{metric}
ds^2=e^{2A(\phi)}(-h(\phi)dt^2+d \vec{x} ^2)+\frac{e^{2B(\phi)}}{h(\phi)}d\phi^2,
\end{eqnarray}
where $A$, $B$, and $h$ are functions of $\phi$, and $\phi$ is also some function of $r$. There is a horizon at $\phi=\phi_H$ which is the solution to the equation $h(\phi)=0$. It is assumed that $A(\phi)$ and $B(\phi)$ are finite at the horizon and the interval $0<\phi<\phi_H$ corresponds to the outside of the horizon. If one works with the Eddington-Finkelstein coordinate system, then the geometry \eqref{metric} is expressed as
 \begin{eqnarray}\label{metric2}
ds^2=e^{2A(\phi)}(-h(\phi)d\tau ^2+d \vec{x} ^2)- 2e^{A(\phi)+B(\phi)} L\,d\tau d\phi.
\end{eqnarray}
%With this form of the metric, the equations of motion take the form
%\begin{subequations}\label{metric2}
%\begin{align}
%&A''(\phi)-A'(\phi)B'(\phi)+\frac{2}{3}=0,\\
%&4A'(\phi)h'(\phi)-B'(\phi)h'(\phi)+h''(\phi)=0,\\
%&\frac{3}{2}A'(\phi)h'(\phi)+h(\phi)(6A'(\phi)^2-1)+2e^{2B(\phi)}L^2V(\phi)=0,\\
%&4A'(\phi)-B'(\phi)-\frac{e^{2B(\phi)}L^2V'(\phi)}{h(\phi)}+\frac{h'(\phi)}{h(\phi)}=0\\
%\end{align}
%\end{subequations}
It will be possible to find a black hole solution if a master function $G(\phi)$, where $G(\phi)=A'(\phi)$, is defined \cite{Gubser:2008ny}. By using this generating function $G(\phi)$ and knowing $V(\phi)$ the different metric components are given by \cite{Attems:2016ugt}
\begin{subequations}\label{metric3}
\begin{align}
A(\phi)&=-\log\left(\frac{\phi}{\phi_0}\right)+\int_{0}^{\phi}d\tilde{\phi}\left(G(\tilde{\phi})+\frac{1}{\tilde{\phi}} \right),  \\
B(\phi)&=\log\left(|G(\phi)|\right) +\int_{0}^{\phi}d\tilde{\phi}\frac{2}{3G(\tilde{\phi})},\\
h(\phi)&=-\frac{e^{2B(\phi)}L^2\left(4V(\phi)+3G(\phi)V'(\phi) \right) }{3G'(\phi)},
 \end{align}
\end{subequations}
where $G(\phi)$ must satisfy the following non-linear master equation
\begin{align}\label{a}
\frac{G'(\phi)}{G(\phi)+\frac{4V(\phi)}{3V'(\phi)}}&=\frac{d}{d\phi}\log\bigg[\frac{1}{3G(\phi)}-2G(\phi) 
+\frac{G'(\phi)}{2G(\phi)}-\frac{G'(\phi)}{2\left(G(\phi)+\frac{4V(\phi)}{3V'(\phi)}\right)}\bigg]. 
\end{align}
%In order to solve \eqref{a}, one needs  two initial conditions. At the horizon $\phi_H$, the condition $h(\phi_H)=0$ along with \eqref{metric} 
%can fix the required initial conditions as follow
%\begin{subequations}\label{metric3}
%\begin{align}
%G(\phi_{H})&=-\frac{4V(\phi_{H})}{3V'(\phi_{H})},\\
%G'(\phi_{H})&=\frac{2}{3}(\frac{V(\phi_{H})V''(\phi_{H})}{V'(\phi_{H})}-1).
% \end{align}
%\end{subequations}
The expression for temperature is finally given by
 \begin{eqnarray}
 \frac{T}{\Lambda}=-\frac{R_{UV}^2V(\phi_H)}{3\pi\phi_H}\exp\{\int_{0}^{\phi_H}d\phi\left(G(\phi)+\frac{1}{\phi}+\frac{2}{3G(\phi)}\right)\}.
 \label{tem}
 \end{eqnarray}
 \section{Review on the entanglement entropy, Complexity and Subregion complexity}
\begin{itemize}
\item \textbf{Entanglement entropy:}  Consider a constant time slice in a $d-$dimensional quantum field theory and divide it into two spatial regions $A$ and $\bar{A}$ where they are complement to each other. The entanglement entropy $S_{A}$ measures how much information is hidden inside $A$  defined as the Von Neumann entropy of the reduced density matrix
\begin{eqnarray}
S_{A}=-tr \rho_{A}\log\rho_{A},
\end{eqnarray}
where $\rho_{A}$ is the reduced density matrix of $A$, given by $\rho_A=Tr_{\bar{A}}$. The AdS/CFT correspondence provides an elegant way to compute the entanglement entropy in terms of a geometrical quantity on the bulk. This so called holographic entanglement entropy formula, first proposed by Ryu and Takayanagi  \cite{Ryu:2006bv,Ryu:2006ef}
\begin{eqnarray}\label{RT}
S_{A}=\frac{Area(\gamma_{A})}{4G_{N}^{d+2}},
\end{eqnarray}
where $S_{A}$ is the holographic entanglement entropy for the subsystem $A$, $\gamma_{A}$  is a codimension-two minimal area surface whose boundary $\partial \gamma_{A}$ coincides with $\partial A$, and $G_{N}^{d+2}$  is the $d+2-$ dimensional Newton constant.
\end{itemize}
\begin{itemize}
\item \textbf{Complexity:}
There are two different proposals for the holographic complexity namely the (CA) conjecture and the (CV) conjecture. The volume complexity prescription states that
\begin{eqnarray}
{\cal{C}}_V=\frac{V}{G_N l} ,
\end{eqnarray}
where $V$ is the volume of the maximal codimension-one bulk time slice, anchored at boundaries at some specific times. $l$ is a length scale associated with the geometry which needs to be chosen for each case at hand and $G_N$ is the Newton's constant. Action proposal states that complexity is proportional to the bulk action evaluated in a certain spacetime region known as the Wheeler-De Witt (WdW) patch 
\begin{eqnarray}
{\cal{C}}_A=\frac{I_{WdW}}{\pi \hbar} .
\end{eqnarray}
The WDW patch can be defined as the domain of dependence of any Cauchy surface on the bulk which asymptotically approaches the time slice $Σ$ on the boundary. The action proposal is more reasonable than the volume one in the sense that there is no need to fix a length scale $l$ by hand.
 \end{itemize}
 \begin{itemize}
\item \textbf{Subregion complexity:}
  The complexity for a subregion A on the boundary equals to the volume of codimension-one Ryu-Takayanagi extremal surface enclosed by  $\gamma_{A}$ which is given by the following form
  \begin{eqnarray}
{\cal{C}}(A)=\frac{\text{Volume}(\gamma_{A})}{8\pi RG_N} ,
\label{CV}
\end{eqnarray}
 where $R$ is AdS radius and $\cal{C}$$(A)$ is known as the holographic subregion complexity for the subregion $A$.
 \end{itemize}
 
\section{Analytical prescription}
In this paper we study the subregion complexity using the volume prescription written in \eqref{CV} for a non-conformal background geometry \eqref{metric}. To get some insight  we take a quick look at the analytical calculations.
 
Consider a general asymptotically AdS$_{d+2}$ metric
 \begin{eqnarray}
ds^2=-f_1(r)dt^2+f_2(r)dr^2+f_3(r)d{\vec{x}}^2,
\label{gamma}
\end{eqnarray}
 where $f_1(r)$, $f_2(r)$ and $f_3(r)$ are some arbitrary functions. The boundary is also located at $r=\infty$. We  consider a strip-like  boundary entangling region $A$ in the $\vec{x}$ directions at a constant time slice. The  entangling region can be parameterized as
\begin{eqnarray}
-\frac{l}{2}\leq x_1\equiv x(r)\leq\frac{l}{2},\qquad -\frac{L}{2}\leq x_i\leq\frac{L}{2},\qquad i=2,.....,d,\qquad L\gg l.
\label{entangling region}
\end{eqnarray} 
Extremal surface is translationally invariant along $x_i$ , $i=2,.....,d$, and the profile of the surface on the bulk is $x(r)$. Area of the surface is given by
\begin{eqnarray}
\text{Area}(\gamma_{A})=L^{d-1}\int dr \; f_3(r)^{\frac{d-1}{2}}\sqrt {f_2(r)+f_3(r)\: x'(r)^2},
\label{gamma1}
\end{eqnarray} 
where $'=\frac{d}{dr}$. Extremizing the area yields the equation of motion for $x(r)$. Since there is no explicit $x(r)$ dependence, the corresponding Hamiltonian is a conserved quantity. The solution for $x(r)$ is then computed as the following one-dimensional integral 
\begin{eqnarray}
x(r)=\int_ {r\ast}^r dr \; \sqrt{\frac{f_3(r\ast)^d f_2(r)}{f_3(r)^{d+1}-f_3(r\ast)^df_3(r)}}.
\label{x(r)}
\end{eqnarray} 
 Using the above expression the relation between $l$ and $r\ast$ is also given by 
\begin{eqnarray}
\frac{l}{2}=\int_ {r\ast}^{r_\infty} dr \; \sqrt{\frac{f_3(r\ast)^d f_2(r)}{f_3(r)^{d+1}-f_3(r\ast)^df_3(r)}},
\label{l(r)}
\end{eqnarray} 
where $r_\infty\equiv r(x=\infty)$ is a cutoff for the  boundary of the underlying geometry introduced to regulate possible divergences arising
due to the UV behavior of the metric.
 Using \eqref{RT}, \eqref{gamma1} and \eqref{x(r)} the holographic entanglement entropy is expressed as
 \begin{eqnarray}
S_{A}(r\ast)=\frac{2L^{d-1}}{4G_{N}}\int_{r\ast}^{r_\infty} dr \; \frac{f_3(r)^{d-\frac{1}{2}}f_2(r)^{\frac{1}{2}}}{\sqrt{f_3(r)^d-f_3(r\ast)^d}}.
\label{gamma}
\end{eqnarray} 
 Due to  UV divergence structure of holographic entanglement entropy we introduce a finite subtracted holographic entanglement entropy called relative holographic entanglement entropy $\tilde{S}$ which has the following form
  \begin{eqnarray}
\hat{S}\equiv\frac{S- S_{AdS}}{S_{AdS}},\label{D}
\end{eqnarray} 
where $S$ and ${S}_{AdS}$  are the holographic entanglement entropy corresponding to non-conformal and AdS geometry, respectively.
Now we would like to compute the volume of the extremal surface stretching inside the region surrounded by the entangling surface. The volume functional is
  \begin{eqnarray}
\text{Volume}(\gamma_{A})=2L^{d-1}\int_{r\ast}^{r_\infty}dr \; x(r)\sqrt{f_2(r)f_3(r)^d}.
\label{V(r)}
\end{eqnarray} 
Using \eqref{CV}, \eqref{x(r)} and \eqref{V(r)} the holographic subregion complexity is written by
 \begin{eqnarray}
{\cal{C}}_A(r\ast)=\frac{2L^2}{4G_{N}}\int_{r\ast}^{r_\infty} dr \;\sqrt{f_2(r)f_3(r)^3} \int_ {r\ast} ^{r} dr \; \sqrt{\frac{f_3(r\ast)^d f_2(r)}{f_3(r)^{d+1}-f_3(r\ast)^df_3(r)}}.
\label{gamma}
\end{eqnarray}
The above holographic subregion complexity is divergent, thus we  define a finite  holographic subregion complexity called relative subregion complexity $\tilde{\cal{C}}$ which is given by
\begin{eqnarray}
\hat{\cal{C}}\equiv\frac{{\cal{C}}- {\cal{C}}_{AdS}}{{\cal{C}}_{AdS}},\label{C}
\end{eqnarray} 
where $\cal{C}$ and ${\cal{C}}_{AdS}$  are the holographic subregion complexity corresponding to non-conformal and AdS geometry, respectively. For convenience, from now on we use the convention that $G_N = 1$  and fix $L=1$.

There are bulk transitions when one would like to compute the behavior of the volume inside the minimal surface. Depending on the ratio $\frac{x}{l}$, there is a transition point  between  connected and disconnected minimal surfaces which we call it disentangling point $x_{DT}$. At the transition point  the connected and disconnected surfaces  have changed their role and therefore one would expect a transition value in the $\tilde{\cal{C}}$.
To study the behavior of the volume inside the minimal surface we  define the renormalized entanglement entropy $\tilde{S}$ and the  renormalized subregion complexity $\tilde{\cal{C}}$ as
\begin{eqnarray}
\tilde{S}\equiv  (S_{conn}-S_{disc}),\qquad
\tilde{\cal{C}}\equiv\frac{{\cal{C}}_{conn} - {\cal{C}}_{disc}}{2l},
\label{C,S}
\end{eqnarray}
where $S_{conn}$, ${\cal{C}}_{conn}$, $S_{disc}$ and ${\cal{C}}_{disc}$ are the entanglement entropy and the subregion complexity of the connected extremal surface and the entanglement entropy and the subregion complexity of the disconnected extremal surface, respectively.
% In the case of the EE of two parallel strips, the disconnected surface is  the addition of two strips of length $l$ and the connected one is composed of two strips of length $x$ and $2l+x$. The volume of the connected surface is calculated by subtracting the volume of a strip of length $x$ from the volume for a strip of length $(2l + x)$ and the volume of the disconnected surface is simply the volume of two strips whose length are $l$.\\
\section{Numerical results}
Having set up the general frame work of  holographic subregion complexity, now we are ready to study
the volume prescription of subregion complexity in the non-conformal model.
\subsection{Zero temperature}
\begin{figure}[h] 
\centering
{ \includegraphics[width=0.32\textwidth]{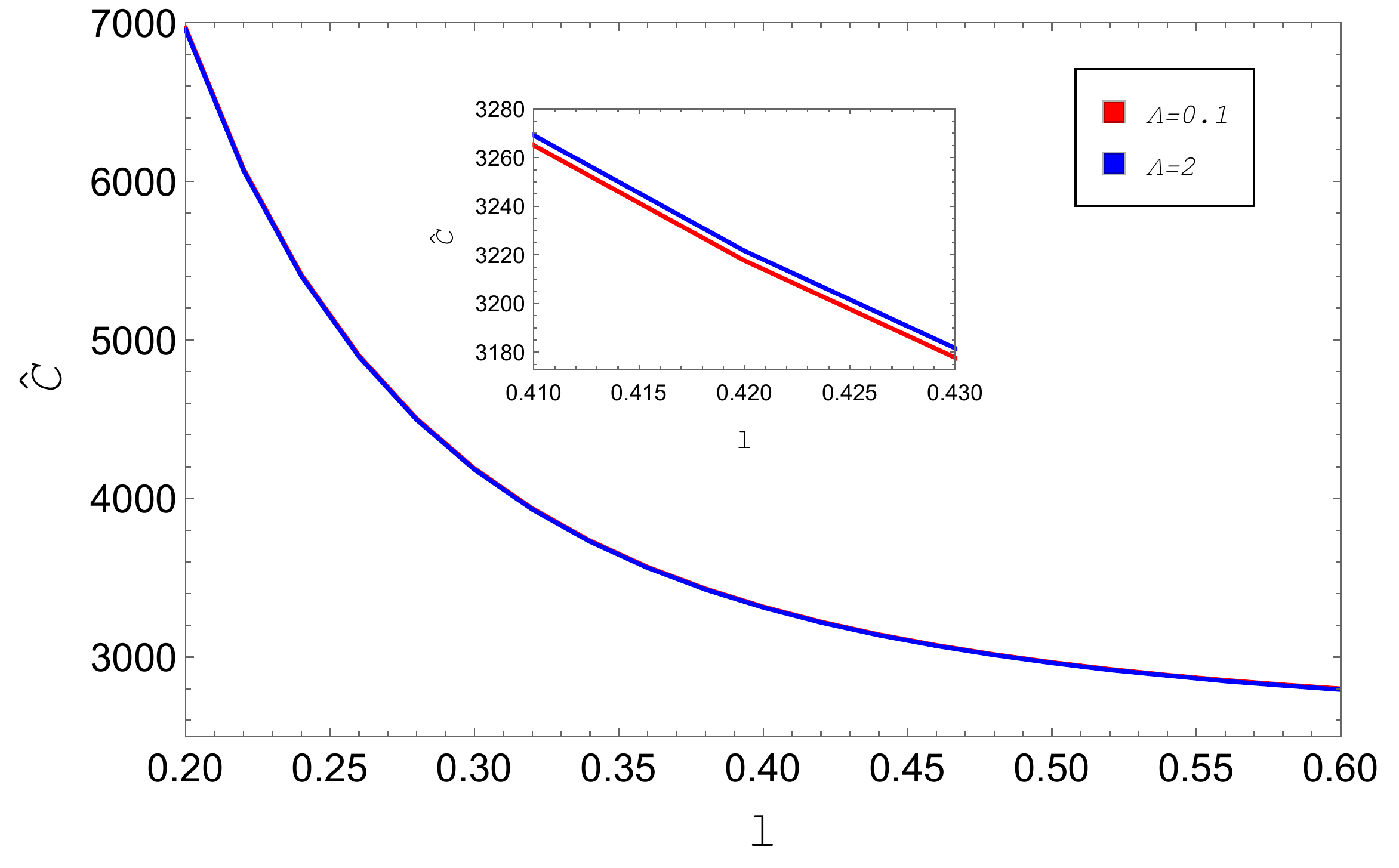} }
{\includegraphics[width=0.32\textwidth]{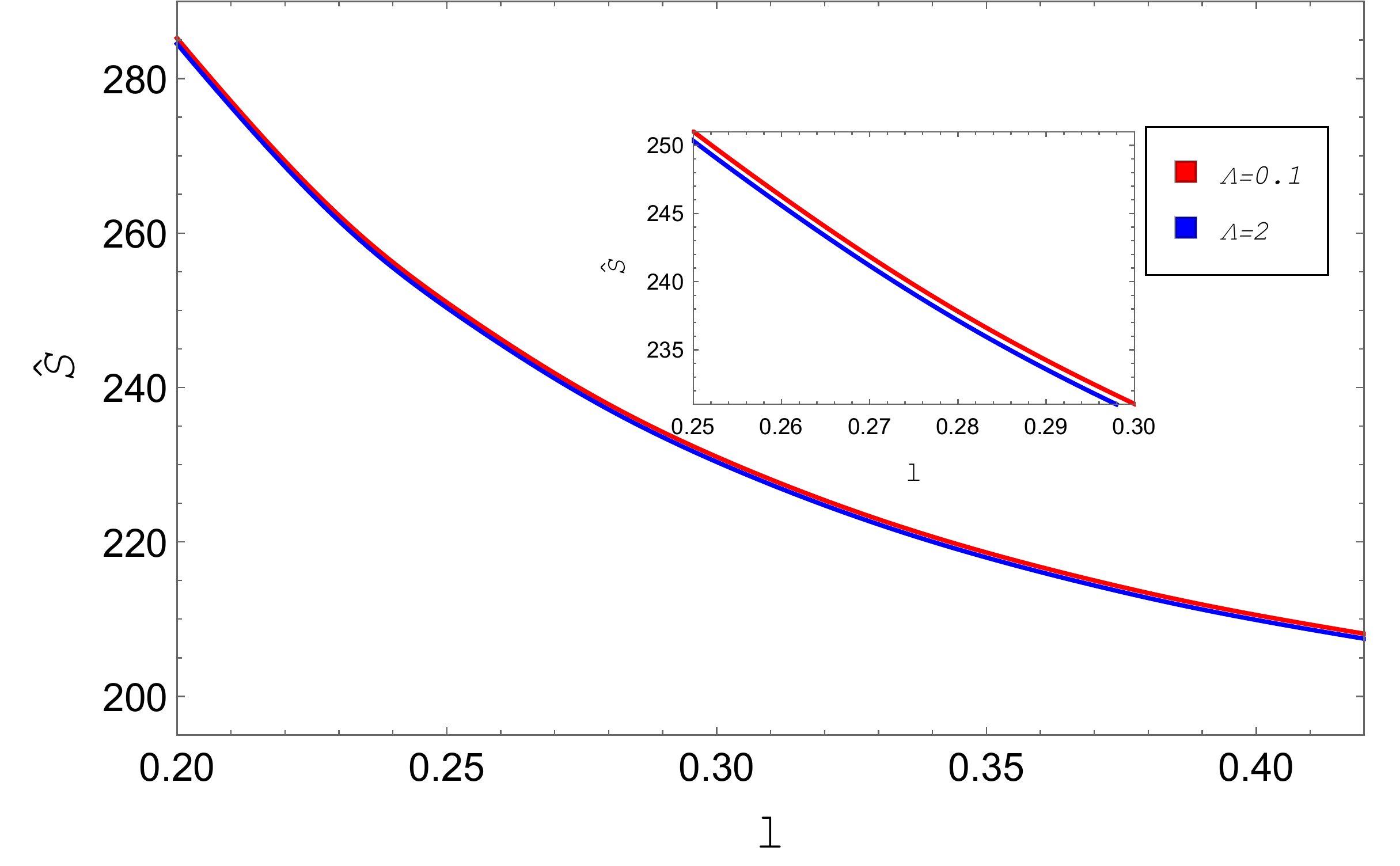} }
{\includegraphics[width=0.32\textwidth]{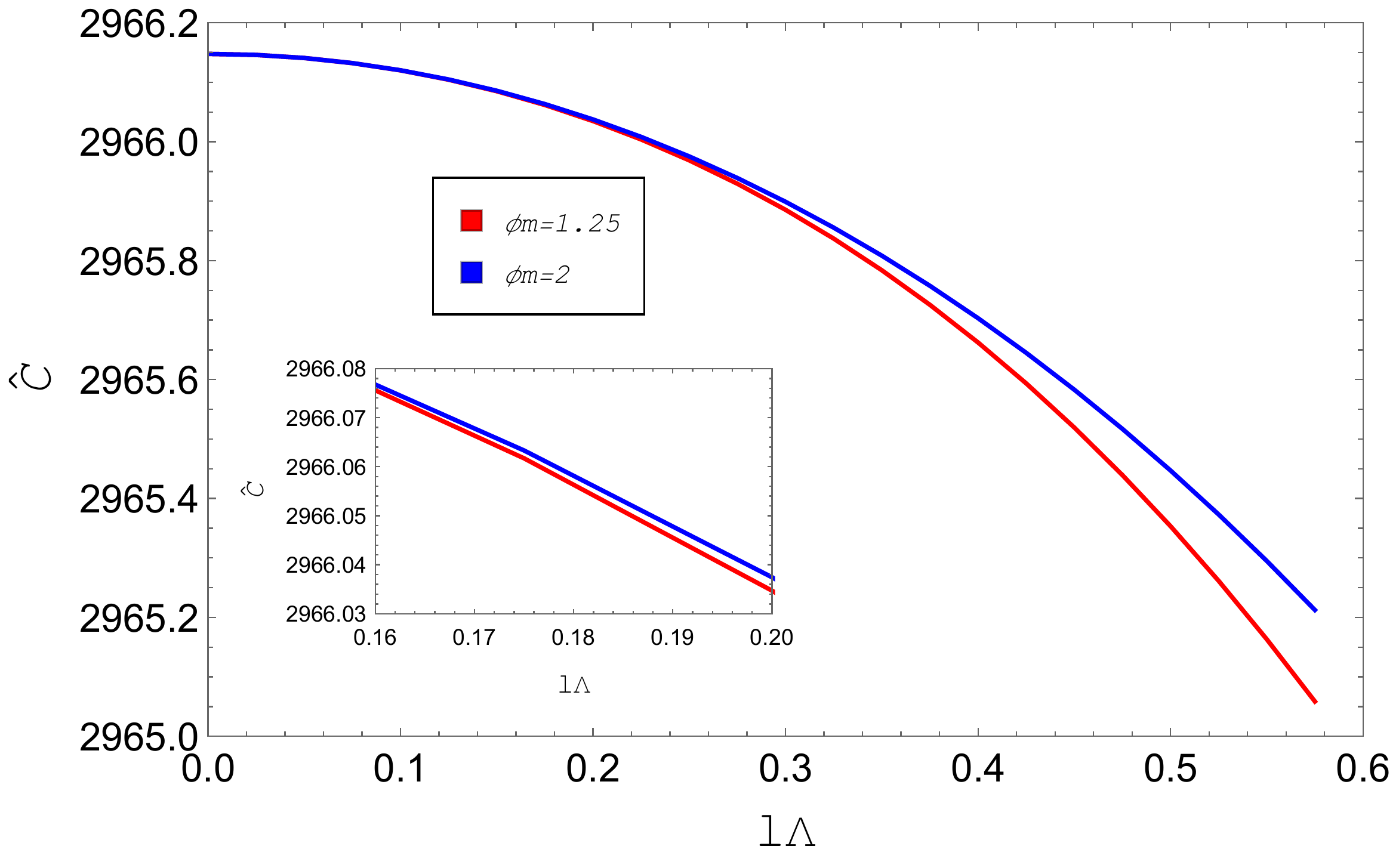} }
\caption{left: The relative subregion complexity $\hat{\cal{C}}$ as a function of length of subregion $l$ for different $\Lambda=0.1$ (red) and $\Lambda=2$ (blue). We have fixed $\phi_M=2$. Middle: The relative entanglement entropy $\hat{S}$ as a function of length of subregion $l$ for different $\Lambda=0.1$ (red) and $\Lambda=2$ (blue). We have fixed $\phi_M=2$. 
Right: The relative subregion complexity $\hat{\cal{C}}$ as a function of  $l\Lambda$ for fixed $l=0.5$. Different curves correspond to distinct $\phi_M=1.25$(red) and $\phi_M=2$(blue).} 
\label{fig1}
\end{figure}
In the left and middle panel of Fig. \ref{fig1} we have plotted the relative the subregion complexity $\hat{\cal{C}}$ and the relative entanglement entropy $\hat{S}$ as a function of length of subregion $l$ for various energy scale $\Lambda$, respectively. 
The common feature is that both quantities are a monotonically decreasing  function of $l$. They firstly start at a positive value, then experience a  significant decreasing and finally reach again to a approximately constant positive value. Indeed, for small enough $l$ relative subregion complexity and relative entanglement entropy  decrease significantly while they do not change for large enough values of $l$.  For large enough values of $l$ one can say that the turning point $z\ast$ approaches AdS$_5$ in the IR limit which is conformal and therefore the $\hat{\cal{C}}$ goes to a constant value.
 Moreover, since the non-conformal relative subregion complexity $\cal{C}$ is larger than the conformal one ${\cal{C}}_{AdS}$ in both UV (small $l$) and IR(large $l$) regime of underlying field theory  then $\hat{\cal{C}}$ is  positive at zero temperature. In other words, by probing the UV and IR regime in field theory  the information needed to prepare the desired state from the reference state in the non-conformal theory is larger than the conformal one. Another point is that increasing $\Lambda$ causes the relative subregion complexity  increases, $i.e$. the larger the energy scale $\Lambda$, the more information required to prepare a final state from an initial state. In contrast, relative entanglement entropy declines if one will increase the energy scale. More correlation between a subregion and its complement, less energy scale.

  In the right panel of Fig. \ref{fig1}, the relative subregion complexity $\hat{\cal{C}}$ with respect to $l\Lambda$ for different model parameter $\phi_M$ has been plotted. The length of subregion $l$ has been fixed.  It is seen that $\hat{\cal{C}}$ experiences two stages as one increases energy scale $\Lambda$, a slowly-reduction stage at small enough $\Lambda$ and then a quickly-decreasing stage at large $\Lambda$ which is in agreement with left panel of Fig. \ref{fig1} except that the rate of decreasing is positive. For small enough $\Lambda$ relative subregion complexity $\hat{\cal{C}}$  is independent of $\Lambda$ compared to the large enough one where it meets relatively moderate change due to the non-conformality. For small enough $\Lambda$ one can deduce that  $\hat{\cal{C}}$ is not a good measure to quantify the deviation from the non-conformality while for large enough one the value of relative subregion complexity depends on $\phi_M$.  As indicated from the behaviour of the $\hat{\cal{C}}$, the larger model parameter $\phi_M$ the larger the deviations from conformality in the theory. Consequently, we can use $\hat{\cal{C}}$ as a good measure of the non-conformality of the theory. It is also obvious that the non-conformal relative subregion complexity decreases by rising $\Lambda$, $i.e$. we need less information to specify the desired state from the initial state in a non-conformal vacuum than the conformal one. Note that there is a slight difference between the two curves at large $\Lambda$. If one, by rising $\phi_M$, increases the difference between the number of degrees of freedom between UV and IR fixed points at large values of $\Lambda$, then relative subregion complexity will grow.\\
\begin{figure}[h] 
\centering
%{\includegraphics[width=0.48\textwidth]{fig55} }
{ \includegraphics[width=0.32\textwidth]{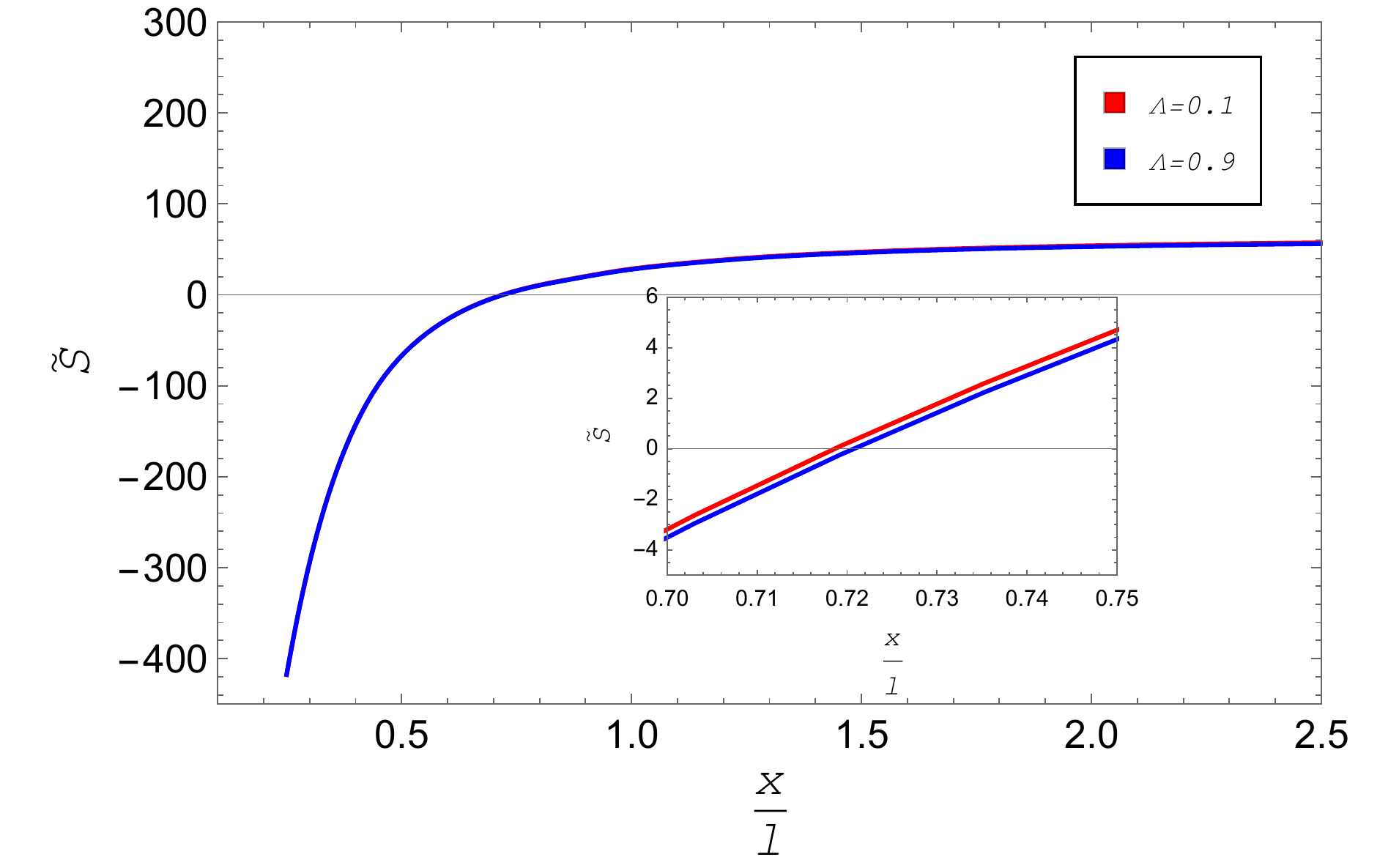} }
{\includegraphics[width=0.32\textwidth]{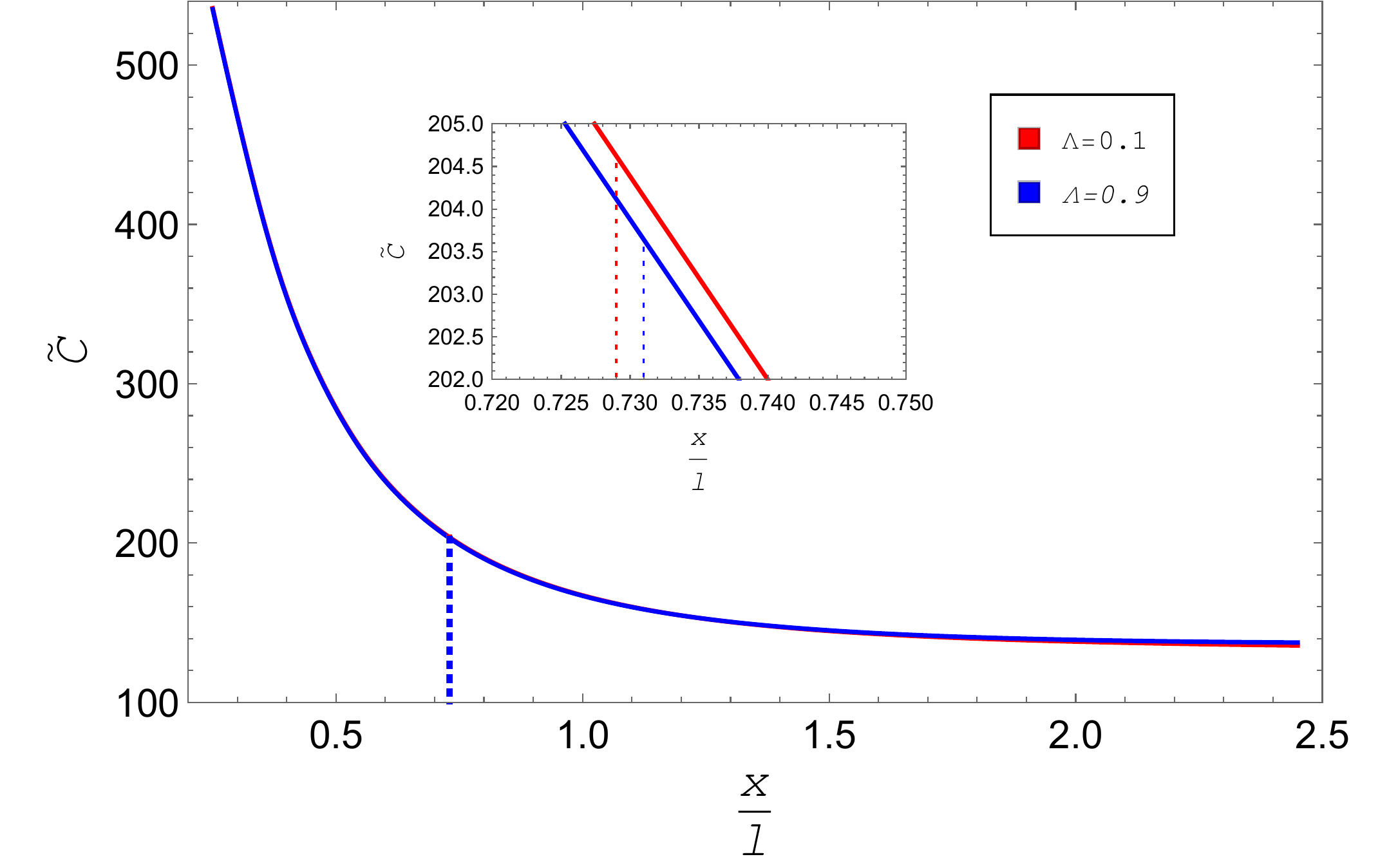} }
{\includegraphics[width=0.32\textwidth]{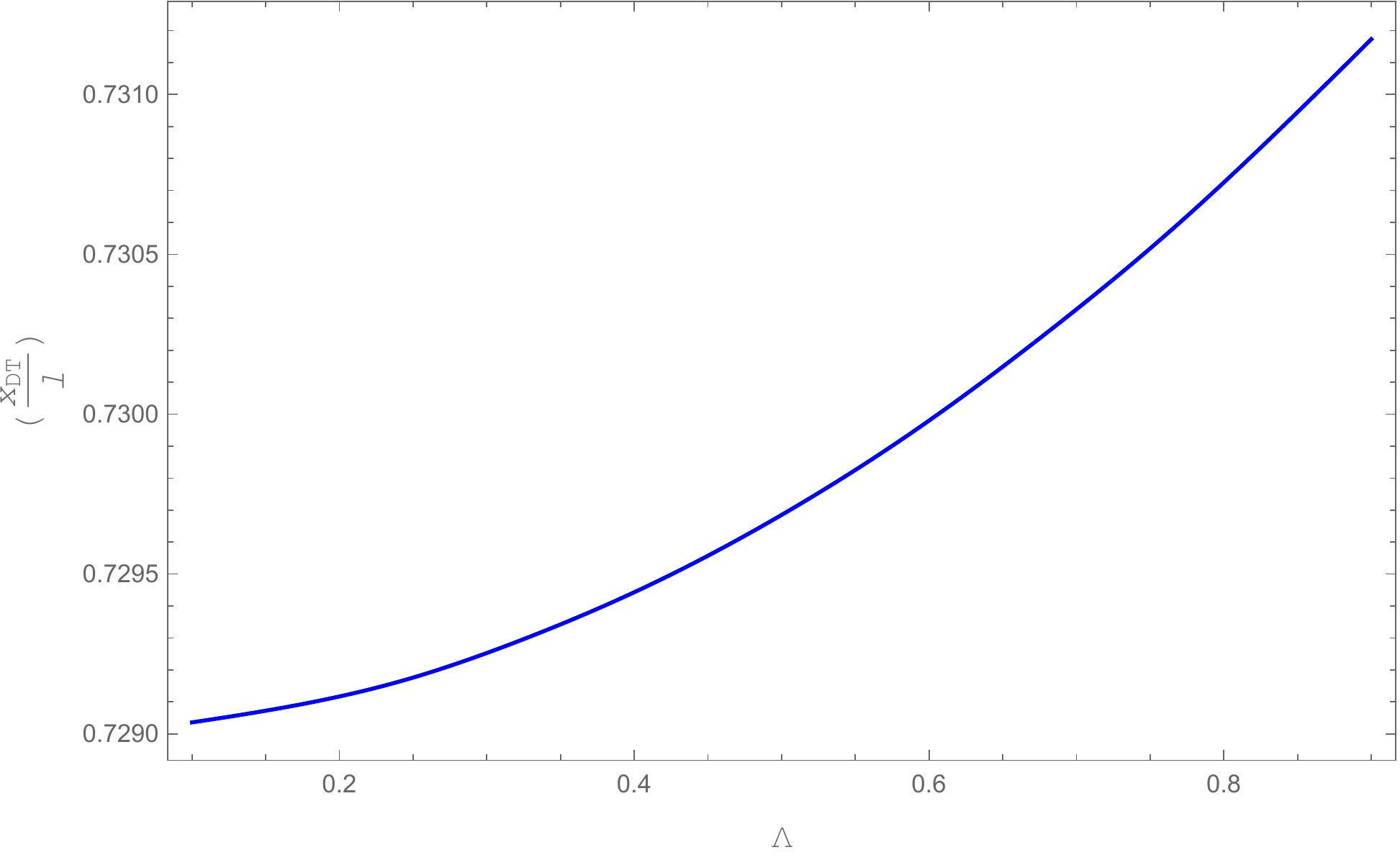} }
\caption{left: The renormalized entanglement entropy $\tilde{S}$ as a function of $\frac{x}{l}$ for different $\Lambda=0.1$ (red) and $\Lambda=0.9$ (blue).\,\, Middle: The renormalized subregion complexity $\tilde{C}$ as a function of $\frac{x}{l}$ for different $\Lambda=0.1$ (red) and $\Lambda=0.9$ (blue) for fixed $\phi_M=2$. \,\,
Right: Rescaled disentangling transition $\frac{x_{DT}}{l}$ in terms of $\Lambda$. We have fixed $\phi_M=2$. In all plots $l=0.1$.} \label{fig4,5}
\end{figure}

In the left and middle panel of Fig. \ref{fig4,5} the renormalized entanglement entropy  $\tilde{S}$ and the  renormalized subregion complexity $\tilde{\cal{C}}$ have been depicted as a function of $\frac{x}{l}$, respectively. We have choosen $l=0.1$ and $\phi_M=2$. It is clearly seen that $\tilde{\cal{C}}$ is always positive, $i.e.$ the disconnected surface is contained inside the connected surface. From the figure, we can see that $\tilde{\cal{C}}$ experiences three stages as $\frac{x}{l}$ is increased. At the first, it decreases quickly, then the decreasing rate drops gradually (a finite jump) around $\frac{x_{DT}}{l}$ and at the end $\tilde{\cal{C}}$ approaches slowly to a saturated value. The underlying finite jump comes  from the disentangling transitions. Note also that, After the transition point, $\tilde{S}$ and $\tilde{\cal{C}}$ do not depend on $x$ and then reach a constant value. Another feature is that by increasing the energy scale $\Lambda$ the renormalized subregion complexity $\tilde{\cal{C}}$ decreases slightly which means the larger $\Lambda$ the smaller $\tilde{\cal{C}}$. Moreover, rising $\Lambda$ causes the disentangling transition  occurs in far distances. Now we would like to analyze relation between the transition point and the energy scale $\Lambda$ which has shown in the right panel of Fig. \ref{fig4,5}. It is observed  that the finite jump in $\tilde{\cal{C}}$ will happen in larger distances if we increase the energy scale. In other words, by increasing $\Lambda$ the connected surface is more dominant than the disconnected one at far separation length between two subregions.\\
 \subsection{Finite temperature}
In  Fig. \ref{fig8,9} we have plotted the relative subregion complexity $\hat{\cal{C}}$, left panel, and the holographic relative entanglement entropy $\hat{S}$, right panel,  as a function of $\frac{\Lambda}{T}$ for  fixed $\Lambda$ and $\phi_M$. Different subregion's length $l$ has been considered.
\begin{figure}[h] 
\centering
{\includegraphics[width=0.48\textwidth]{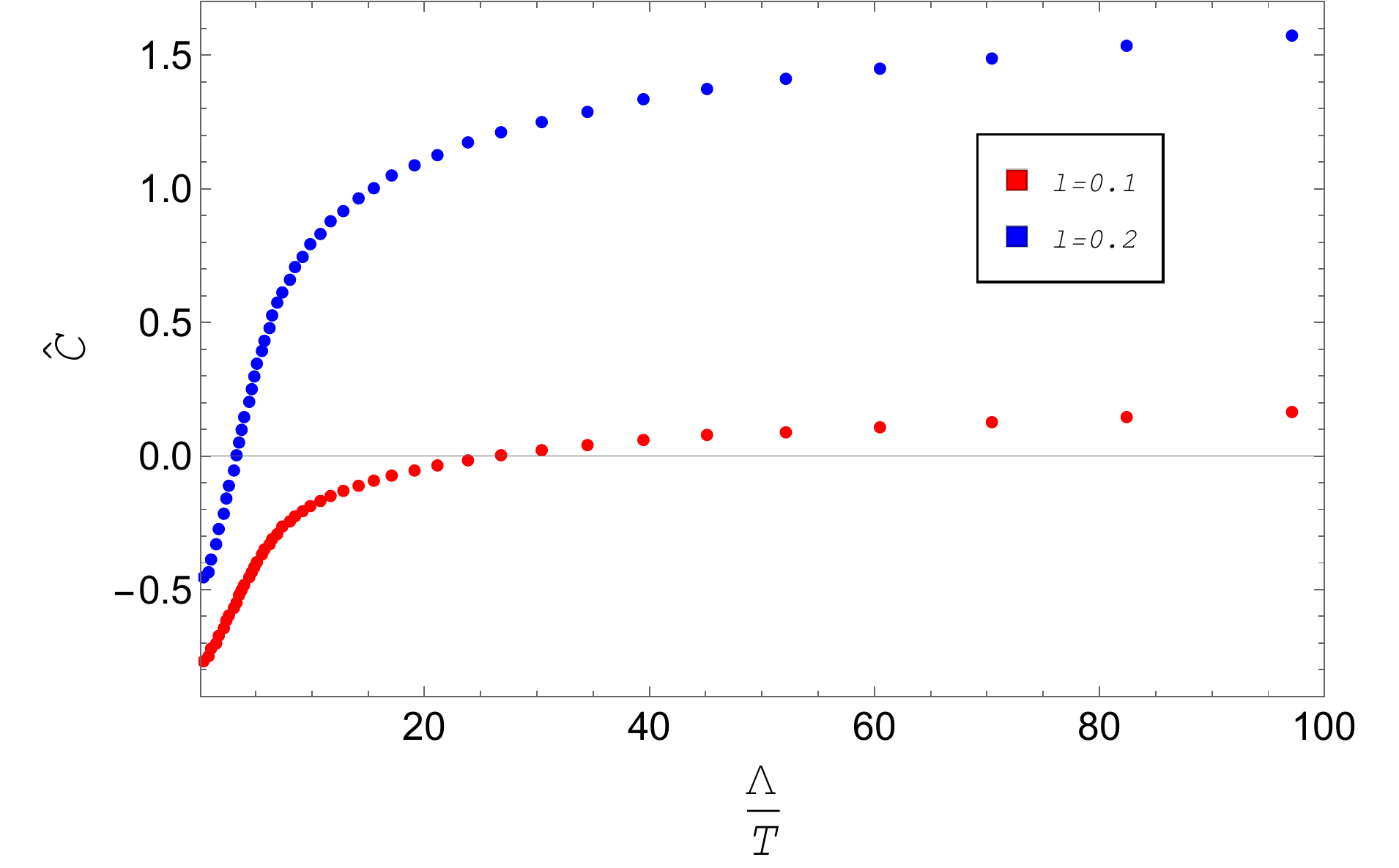} }
{ \includegraphics[width=0.48\textwidth]{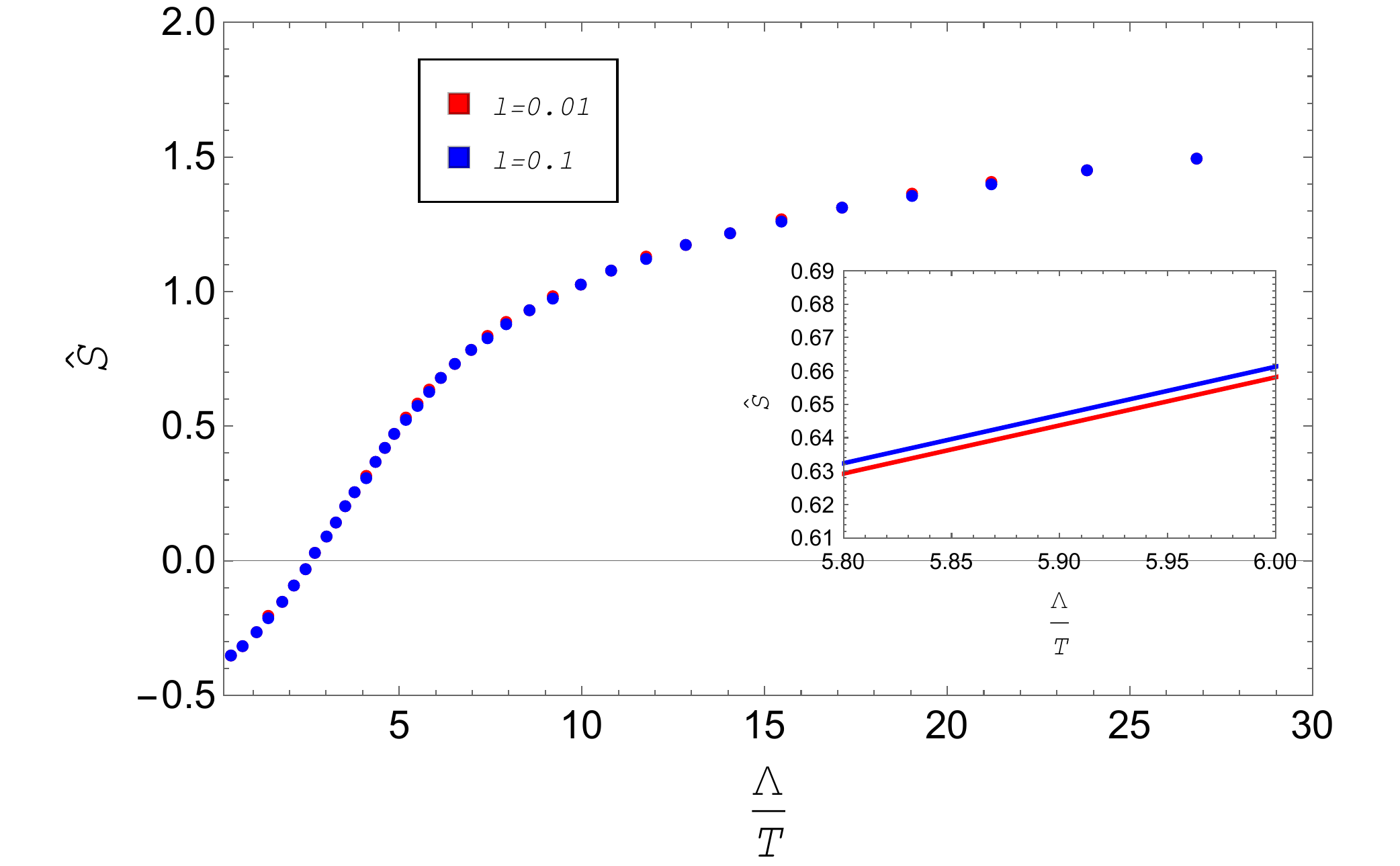} }
\caption{Left: \,The relative subregion complexity $\hat{\cal{C}}=\frac{C(T)-C_{AdS}(T=0.8)}{C_{AdS}(T=0.8)}$ as a function of  $\frac{\Lambda}{T}$ for fixed $\Lambda=0.8$ and $\phi_M=10$. Different curves correspond to different subregion length, $l=0.1$ (red) and $l=0.2$ (blue). Right: The holographic relative entanglement entropy $\hat{S}$ as a function of $\frac{\Lambda}{T}$  for different $l=0.01$ (red) and $l=0.1$ (blue). we have fixed $\Lambda=0.8$ and $\phi_M=10$. } 
\label{fig8,9}
\end{figure}

The common feature is that the holographic relative subregion complexity $\hat{\cal{C}}$ and the holographic relative entanglement entropy $\hat{S}$ will increase if one increases the length of subregion $l$. Unlike $\hat{S}$, the effect of rising $l$ can be significantly seen in the behaviour of $\hat{\cal{C}}$. Another common feature is that both quantities are a monotonically increasing function with the same behaviour at high temperature ($T\gg\Lambda$), intermediate temperature ($T\sim \Lambda$) and low temperature ($T\ll\Lambda$) regime. At low temperature limit where $r_{\ast}\gg r_H$ the leading contributions to the  $\hat{\cal{C}}$ come from the boundary which is asymptotically AdS$_5$. Consequently we should expect that zero temperature $\hat{\cal{C}}$ as the leading term which is seen from both plots. Finite temperature corrections stand for the bulk deviation from  asymptotically AdS$_5$. On the other side, at high temperature the extremal surface tends to wrap a part of the surface and  the full bulk geometry contributes. Hence, the deviation from boundary is tangible as one can see from the above plots. It is also seen that the non-conformal quantities are smaller than the conformal one at high temperatures, $i.e$. there is less correlation between a subregion and its complement and less information needed to prepare a final state from a initial state in the non-conformal model relative to conformal one as one goes to high temperature regime. While, the role is changed at intermediate and low temperatures.
Interestingly, in the left panel, there is a specific temperature denoted by $T^\ast$ where the difference between non-conformal subregion complexity and conformal subregion complexity is equal to zero, $i.e$.  $\hat{\cal{C}}=0$. The point is that by increasing the subregion's length $l$ the corresponding temperature $T^\ast$ increases that is $T^\ast_{l=0.1}<T^\ast_{l=0.2}$. In other words, conformal subregion complexity ${\cal{C}}_{AdS}$ begin to overtake conformal one $\cal{C}$ at higher temperature. If one defines  $\Delta \hat{\cal{C}}=\hat{\cal{C}}_l - \hat{\cal{C}}_{l'} $ as the difference between relative subregion complexity of two subregions whose lengths are $l$ and $l'$, provided that $l'>l$, then it experiences a significant increasing from high to low temperature limit and reaches a constant value at very low temperature at the end. As a result, decreasing temperature has an increasing effect on $\Delta \hat{\cal{C}}$.\\
\begin{figure}[h] 
\centering
{\includegraphics[width=0.48\textwidth]{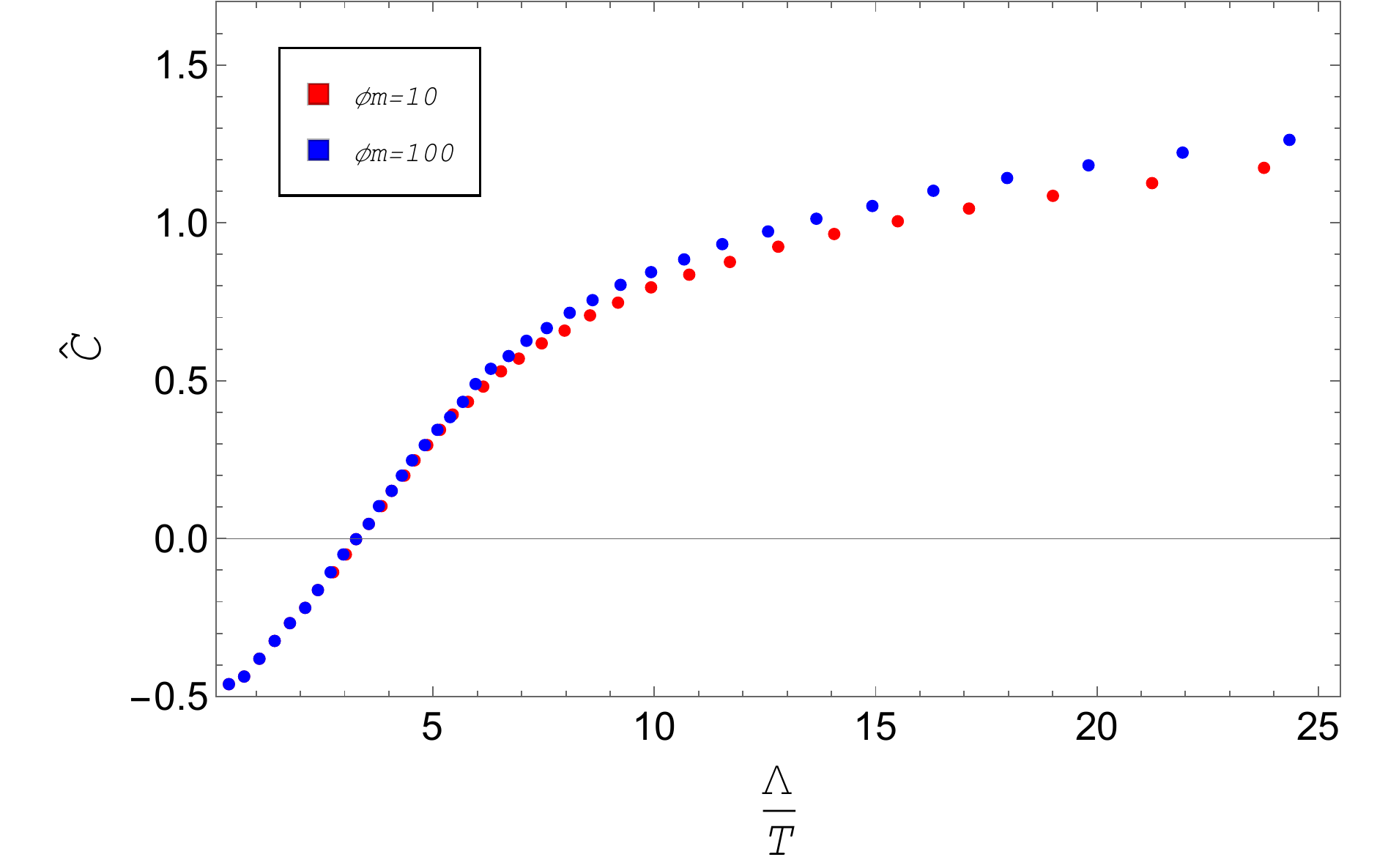} }
{\includegraphics[width=0.48\textwidth]{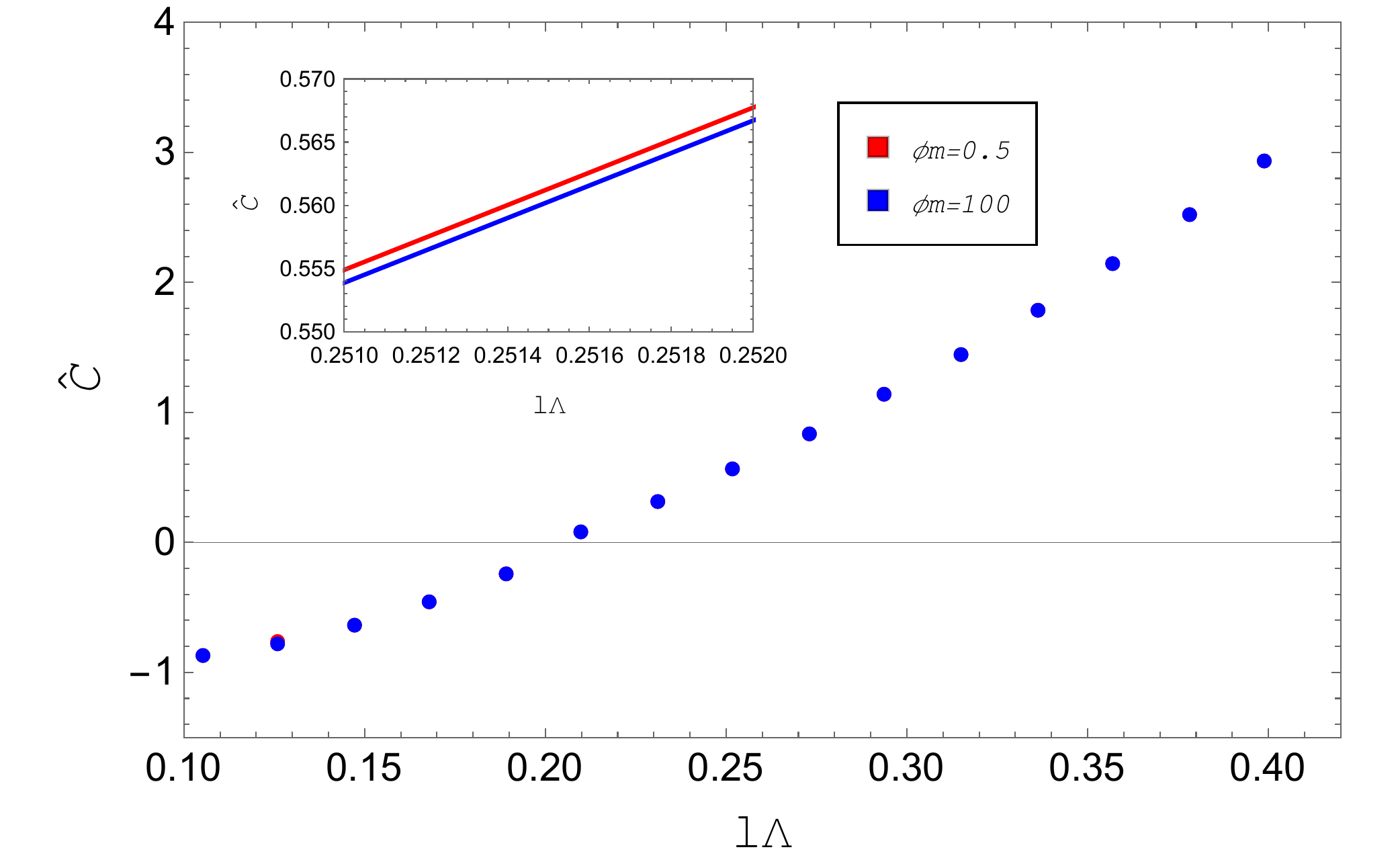} }
\caption{Left: \,The relative subregion complexity $\hat{\cal{C}}=\frac{{\cal{C}}(T)- {\cal{C}}_{AdS}(T=0.8)}{{\cal{C}}_{AdS}(T=0.8)}$ as a function of  $\frac{\Lambda}{T}$ for fixed $\Lambda=0.8$ and $l=0.21$. Different curves correspond to distinct $\phi_M=10$(red) and $\phi_M=100$(blue). Right: \,The relative subregion complexity $\hat{\cal{C}}=\frac{{\cal{C}}(T)- {\cal{C}}_{AdS}(T=0.8)}{{\cal{C}}_{AdS}(T=0.8)}$ as a function of  $l\Lambda$ for fixed $l=0.21$. Different curves correspond to distinct $\phi_M=0.5$(red) and $\phi_M=100$(blue).} 
\label{fig7,10}
\end{figure}

We have  illustrated  the relative subregion complexity $\hat{\cal{C}}$ in terms  of $\frac{\Lambda}{T}$ for fixed subregion's length $l$ and energy scale $\Lambda$  in left panel of Fig. \ref{fig7,10}. Two different model parameter $\phi_M$ have been considered. In the right panel  the relative subregion complexity $\hat{\cal{C}}$ has been plotted as a function $l\Lambda$ for fixed $l$ and different $\phi_M$. We have listed the following results.
\begin{itemize}
\item Left panel

It is seen that the relative subregion complexity is a monotonically increasing function starting from negative values at high and intermediate temperatures, passing through zero value and finally reaches to the positive values at low temperature. The functional behaviour of $\hat{\cal{C}}$ in terms of $\frac{\Lambda}{T}$ is the same as the holographic entanglement entropy which is reported in \cite{Rahimi:2016bbv}. Interestingly, at intermediate and low temperatures, just like zero temperature,  the relative subregion complexity is $\phi_M$ dependent and then we can quantify the non-conformal behaviour of the dual theory. However, such behaviour is not tangibly observed at high temperatures. 
The interesting features corresponding to Fig. \ref{fig7,10} have been pointed in the following.
\begin{itemize}
\item High temperature regime ($T \gg \Lambda$) : The important feature is that non-conformal subregion complexity is smaller than conformal one, $i.e.$ $\hat{\cal{C}}<0$ and then less information is needed to prepare a desired state from a refrence state  for a non-conformal model. Another feature is that $\hat{\cal{C}}$ is independent of model parameter $\phi_M$.  This is due to the fact that the value of the scalar field at horizon is small and hence the physics is sensitive only to the small field behaviour of the scalar potential which is independent of $\phi_M$.
\item Low temperature regime ($T \ll \Lambda$) : The interesting point is that non-conformal subregion complexity is bigger than conformal one, $i.e.$ $\hat{\cal{C}}>0$ which is in complete agreement with Fig. \ref{fig4,5}. At very low temperature, the trend appears to be stable. On the gravity side, the geometry approaches AdS$_5$ and then it seems that  $\hat{\cal{C}}$ does not change significantly. Another point is that the value of  relative subregion complexity is non-zero and depends on $\phi_M$. The larger $\phi_M$ the larger the deviations from conformality in relative subregion complexity. Consequently, we can use $\hat{\cal{C}}$ as a measure of the non-conformality of the theory. 
\item There is a point denoted by $(\frac{\Lambda}{T})^\ast$ where the non-conformal subregion complexity is equal to the conformal one and hence $\hat{\cal{C}}=0$. It is obvious that for  $(\frac{\Lambda}{T})<(\frac{\Lambda}{T})^\ast$, $\hat{\cal{C}}$ is negative and for $(\frac{\Lambda}{T})>(\frac{\Lambda}{T})^\ast$, $\hat{\cal{C}}$ is positive. The interesting feature is that by increasing the model parameter $\phi_M$ the point $(\frac{\Lambda}{T})^\ast$ increases, $i.e.$ $(\frac{\Lambda}{T})^\ast_{\phi_M=10}<(\frac{\Lambda}{T})^\ast_{\phi_M=100}$. In other words, if one increases the difference in degrees of freedom between the UV and the IR fixed points, by rising $\phi_M$, then $T^\ast_{\phi_M=10}>T^\ast_{\phi_M=100}$ and indeed we can say that the deviation from the conformality appears at higher temperature. 
\end{itemize}
\item Right panel

It is  seen that by increasing energy scale $\Lambda$ the relative subregion complexity $\hat{\cal{C}}$ increases starting from negative values and ending to positive values. Note that this behaviour is in complete contrast with zero temperature case which is decreasing function. For small  energy scale $\Lambda$ less work is needed to reach a final state from initial state in non-conformal model than conformal one. While, for large $\Lambda$ the opposite is true. Since the temperature is directly proportional  to the energy scale with a minus sign, according to \eqref{tem}, then by increasing $\Lambda$ one would expect a decreasing temperature. Therefore, by rising $\Lambda$ we approach the low and zero temperature values of $\hat{\cal{C}}$ which are positive and coincide with the results in Fig. \ref{fig1}. In contrast with zero temperature regime, if one raises  the difference in degrees of freedom between the UV and the IR fixed points, by increasing $\phi_M$, the relative subregion complexity becomes smaller. 
\end{itemize}

\subsection*{Acknowledgements}
M. A. would like to  kindly thank M. Ali-Akbari and M. Lezgi  for useful comments and discussions on related topics.

\end{document}